\documentstyle[12pt,aasms4]{article}

\slugcomment{Submitted to {\em Astrophysical Journal Letters}}

\begin{document}

\title{Log-Normal Distributions in Gamma-Ray Burst Time Histories}

\author{Hui Li \& Edward E. Fenimore}
\affil{D436, Los Alamos National Laboratory, Los Alamos, NM 87545}
\affil{hli@lanl.gov}

\begin{abstract}
We propose a new, simple but powerful algorithm to analyze the 
gamma-ray burst temporal structures based on identifying 
non-statistical variations (``peaks'') in the time histories.
Detailed analyses of the bursts from the third BATSE catalog
show that $\sim 30$ bursts have more than 20 peaks individually.
Upon identifying most of the peaks in those bursts,
we show that the peak fluence $S_i$ and peak interval $\delta_i$
distributions within each burst are consistent with log-normal 
distributions. Furthermore, we show that Gaussian (in linear space) 
and power-law distributions for peak fluences are ruled out, as is
the Poisson distribution for peak intervals.

\end{abstract}


\section{Introduction}

Though the origin of Gamma-ray Bursts (GRBs) remains elusive,
the abundant data collected by BATSE on {\it Compton} Gamma-ray
Observatory (CGRO) have provided many constraints on the physical 
modeling of these events. Thanks to BATSE's high temporal resolution
and high signal-to-noise ratio, temporal analyses of bursts have began
to bear fruit. Notably, the bi-modal distribution in the burst durations
(\cite{kle92};\cite{k93}); the claimed time-dilation of peak width 
between strong and weak bursts (\cite{n94} although see
\cite{m94}); and that the peak width narrows
with energy following a power-law (\cite{fen95}). Though
the exact physics is yet unknown to satisfy all these observations,
they are important probes into the burst emission processes.

Here in this {\it Letter}, we propose a new, simple but powerful
peak-finding-algorithm (PFA) that can identify the ``peaks'' (or 
non-statistical variations) within a GRB.  Using this PFA,
we uncover that the distributions of peak fluences and intervals within
each burst are log-normal and several tests are performed to confirm our
findings.

\section{The Peak-Finding Algorithm}
\label{pca-sec}

GRB time histories show vast diversities and Figure \ref{2grb-fig}
shows portions of two complex bursts: trigger 678 and 1606.
Norris et al. (1996) have used a pulse-fitting
algorithm based on $\chi^2$-fitting of individual 
pulses. However, their method fails to converge when the burst is 
very complex, particularly for the two bursts shown here.

The detailed procedure to find all the peaks and valleys in a burst
is as follows: (a) First, we fit the burst background using a 
linear function $B(t)$ to the pre- and post-burst regions.
(b) During the burst, every count bin that has more counts 
than the neighboring bins (both sides) is a candidate peak with 
count $C_p$ at time $t_p$.
(c) We then search on both sides of each candidate peak for counts 
$C_1$ (at $t_1$) and $C_2$ (at $t_2$) so that the conditions
$C_p - C_{1,2} \geq N_{\rm var} \sqrt{C_p}$ are satisfied.
(d) The search will stop either when both $C_1$ and $C_2$ are found,
in which case $C_p$ becomes a ``true'' peak, or when counts higher
than $C_p$ (on either side of $t_p$) is encountered, in which case
$C_p$ is not a true peak and discarded. After this step, all the
peaks ($N_k$ in total) should have been identified. 
(e) We then locate the minima between two successive peaks as valleys. 
Note that $C_1$ and $C_2$ are not necessarily the valleys. 
The valley at the beginning (end) of a burst is chosen
from the location where counts start deviating from (dimming into)
the background.

The above procedure has been implemented and
in Figure \ref{2grb-fig}, we present the selected peaks (up-triangles)
and valleys (down-triangles) for burst 678 and 1606. Note that only
a small portion of the time histories is shown for each burst and
there 58 and 35 peaks in total (with $N_{\rm var}=5$) for
burst 678 and 1606 respectively.

Assume that the peak is $C_p(t_p)$ and two neighboring valleys 
are $C_1(t_1)$ and $C_2(t_2)$ after applying the 
peak-finding algorithm, as schematically shown in Figure \ref{pfa-fig}
along with the background level $B(t)$.
The interval between adjacent peaks, or the waiting time
between successive peaks, is $\delta_i = t_{p_{i+1}} - t_{p_i}$ with
$i = 1, \dots, N_k -1$; and the count fluence within the $i$th peak
is defined as $S_i = \sum_{t_1}^{t_2} \left [ C(t) - B(t) \right]$.
Note that we use count fluence rather than photon energy fluence,
though they are simply related by the mean photon energy
if it remains approximately the same throughout the burst. 
So, there are $N_k-1$ intervals but $N_k$ peak fluences in each burst.

The number and position of peaks within a GRB, as one might suspect, 
depends somewhat on the exact value of $N_{\rm var}$, but the effect
is small when $3 \leq N_{\rm var} \leq 5$. 
In this {\it Letter}, we choose $N_{\rm var}$ to be 5. 
Tests using various $N_{\rm var}$ show that our 
conclusions are not affected by this choice.

\section{Log-Normal Distributions of Peak Fluences and Intervals}

A two-parameter log-normal distribution is represented 
as follows (\cite{ab57}; \cite{cs88}):

\begin{equation}
\label{lnd-eq}
f(x)= \left \{ \begin{array}{ll}
{1 \over {\sqrt{2\pi} \sigma}}~
\exp^{-(\log x-\mu)^2/2\sigma^{2}} & x > 0 \\
0 & x \leq 0
\end{array}
\right.
\end{equation}

\noindent where $f(x)$ is the probability density function for 
$\log x$, $\mu$ and $\sigma^2$ are the two parameters corresponding to
the sample mean and variance of $\{\log x_i\}$. 
The cumulative log-normal distribution is ${1\over 2}~{\rm erfc}(y)$,
where ${\rm erfc}(y)$ is the complementary error function and
$y = (\mu - \log x_i)/\sqrt{2\sigma^2}$. 

Utilizing the BATSE DISCSC data (64 ms temporal resolution and 
4-channel spectral resolution), we analyze most of the bursts in
the third BATSE catalog (\cite{mee96}) and identify their peaks
using our PFA with $N_{\rm var} = 5$. 
Since we require that each burst must have more than 20 peaks 
so that a distribution can be built, and all the peaks must 
be 5 sigma above background, 32 bursts are selected that meet the
criteria. The requirement of more than 20 peaks in each burst is somewhat
arbitrary and it is mainly due to the concern that the statistical
significance on fewer than 20 points in a distribution is, 
in general, questionable.
We have done tests by relaxing this requirement and found
that our conclusions remain intact. 

All 32 selected bursts tend to be bright, long, complex and, 
on average, each of them has $\sim 35$ peaks. In the following,
we concentrate on the peak fluences $S_i$ and peak intervals 
$\delta_i$ distributions. For a burst with $N$ peaks, we define 
two sample means for fluences: $\mu_{ln} = \sum^{N}_1 \log(S_{i}) / N$ and 
$\mu_{n} = \sum^{N}_1 S_{i} / N$; and two sample variances:
$\sigma^2_{ln} = \sum^{N}_1 (\log(S_{i}) - \mu_{ln})^2 / (N-1)$ and
$\sigma^2_{N} = \sum^{N}_1 (S_{i} - \mu_{n})^2 / (N-1)$. Here,
the subscript $ln$ represents $\log$ and $n$ for linear. 
Sample means and variances for $\{\delta_i\}$ can be defined similarly.

We find that, for all 32 bursts, 
the distributions of the peak fluences $\{S_i\}$ and the peak 
intervals $\{\delta_i\}$ in {\it individual bursts} are consistent 
with log-normal distributions (i.e. the differential 
distributions of $\{\log(S_i)\}$ and 
$\{\log(\delta_i)\}$ are consistent with Gaussian). 
We reach this conclusion by performing $\chi^2$-fitting. 
The peak fluences and intervals from each bursts are binned (5 bins) 
and fitted with log-normal distributions with two parameters 
(mean and variance). In Figure \ref{chi2prob-fig}$a$ and $b$, we
give the $\chi^2$-fitting probabilities for 
$\{S_i\}$ and $\{\delta_i\}$ for each burst, respectively.
Note that the probabilities are reasonably uniformly distributed
between 0 and 1, implying that log-normal is, in general, an
acceptable hypothesis. However, we caution that $\chi^2$-fitting
might not be an accurate measure of statistics here due to the small 
number of points in each bin though they do satisfy the limit
given by Lindgren (1976) that the sample size should be roughly
4 or 5 times the number of bins. 

Perhaps the peak fluence distribution is similar to log-normal,
but instead follows some other distributions, such as a truncated
power-law. We have performed fitting to peak fluences assuming
they are distributed as power-laws (again with two parameters, 
the slope of the power-law and the lower cutoff). Their probabilities are
shown in Figure \ref{chi2prob-fig}$c$ (notice that probability is now 
in logarithmic scale).  More than half of the bursts
have probabilities less than $10^{-3}$ and the acceptable
probabilities for some bursts may just be due to their small number of 
peaks (note the trend for probability becoming exceedingly smaller 
when peak number increases).

Even though peak fluence and peak interval distributions 
in individual burst are consistent with log-normal distribution, 
their confidence is nevertheless affected by the small number of peaks. 
A more stringent test on peak fluence distribution can be performed
as follows: for each burst, we rescale each $S_i$ so that
the new $\mu_{ln} = \mu_{n} = 0$ and $\sigma^2_{ln} = \sigma^2_{n}
= 1$. As a result, all 32 bursts now have the same $\mu$($=0$)
and $\sigma^2$($=1$) in the peak fluence distribution.
We then put all the peaks from all 32 bursts together and sort them in the
ascending order for the {\it scaled} $S_i$.
This combination of all peaks from all bursts after 
the rescaling (1107 peaks total) removes the uncertainty in small
number of peaks. In Figure \ref{sumks-fig}$a$ and $b$,
we present the cumulative distribution of rescaled 
$\log S_i$ and $S_i$ as compared to the cumulative 
Gaussian distribution with mean $=0$ and variance 
$=(N_{\rm pk} - N_{\rm bur}) /(N_{\rm pk} - 1) \approx 0.97$, 
where $N_{\rm pk}=1107, N_{\rm bur}=32$ are the total number 
of peaks and bursts, respectively. 
This extremely good fit in Figure \ref{sumks-fig}$a$ strongly 
indicates that the peak fluences in GRBs are distributed 
log-normally rather than normally (Figure \ref{sumks-fig}$b$).

The same procedure can be applied to the peak interval distribution
and the results for $\log\delta_i$ are shown in Figure 
\ref{sumks-fig}$c$, it is well fitted by the log-normal distribution.
Special attention should be taken, however, in making the
comparison with the hypothesis that peak intervals are distributed
{\it randomly}. The probability for having another peak after time
$\delta_i$ is $e^{-\delta_i / <\delta_i>}$ (i.e. the Poisson probability
for zero event), and the corresponding cumulative distribution
is $1 - e^{-x}$, where $0 \leq x < \infty$. Therefore, we rescale the
peak intervals in each burst by its mean $<\delta_i>$ and put all
the peaks from all bursts together and plot it against 
$1 - e^{-x}$. This is shown in Figure \ref{sumks-fig}$d$.
The inconsistency is obvious, therefore, the hypothesis that 
peaks appear randomly can be ruled out.

Though the fitting to log-normal distribution is extremely good, 
small deviations in Figure \ref{sumks-fig}$a$ and 
\ref{sumks-fig}c are still visible. Possible explanations include
the following: For peak fluence, our
definition of $S_i$ (the area enclosed by two successive valleys)
may be a slight underestimation, especially for large peaks due to
the fluences lost in the tails which extend into the neighboring
peaks. However, this effect is mutual for adjacent peaks so the net
difference can be estimated to be very small (a few percent at most).
This is enough to account for the tiny asymmetry (bright end)
in peak fluence distribution.
The small asymmetry in peak interval distribution, we believe, 
is caused by the limited temporal resolution (64ms) of the data. 
In fact, the peak interval has to be larger than a few bins 
(by definition) whereas equation \ref{lnd-eq} assumes $x$ can be
arbitrarily close to 0. More precisely, a three-parameter log-normal 
distribution (\cite{ab57}) can be employed but we omit any
detailed comparison here as the real correction can only come from
analyzing the high temporal resolution data (such as Time-Tagged Events
(TTE) data of BATSE). 

McBreen et al. (1994) have also suggested that the intervals between
peaks in GRBs may be distributed log-normally, based on their analyses
of 34 long, intense KONUS bursts with 0.25 s time resolution
(1 s resolution after 34 s in a burst).  
Their analyses and results, however, are very different from ours 
in several aspects. By using a rigorous peak selection process, 
we have identified individual entities within the GRB time history that 
are distinct, and found that each {\em individual} burst has a 
log-normal distribution.
In contrast to our $\sim 35 $ peaks per bursts, McBreen et al. (1994) had
only 4-5 ``peaks'' on average.  More questionably, peak intervals from
different bursts with different brightnesses were put together to form one
distribution with no normalization.
We find that great caution has to be taken
when putting together quantities from different bursts due to the bias 
that fewer peaks on average will be selected from weaker bursts. 
No alternative, non-lognormal distributions were considered in
their analyses either.

\section{Summary}

Our findings that peak fluences and peak intervals in individual
GRB are distributed log-normally probably have far-reaching 
implications on the production of GRBs and consequently their origin,
though it is beyond the scope of this {\it Letter}.
The non-random appearance of peaks seems directly pointing
to a central engine, and interestingly, similar behavior has 
also been found in the interval (waiting time) distribution for the
successive events from the soft gamma-ray repeaters (SGRs)
(\cite{hur94}), which have been suggested to come
from the neutron star quakes (\cite{cheng96}). 
Further analyses on the statistical properties of GRB time
histories, such as the detailed comparisons with those
found in SGRs, will be presented in the future publications.

\acknowledgments

Useful discussions with B. Cheng, S. Colgate and R. Epstein 
are appreciated.  H.L. gratefully acknowledges the support 
by the Director's Postdoc Fellowship at Los Alamos National Lab. 
This work is partially supported by the GRO Guest investigator program.

\clearpage

\begin{figure}
\caption{Portions of time histories for 2 complex
GRBs $678$ and $1606$. Filled up-triangles show
chosen peaks and down-triangles show valleys using our 
peak-finding algorithm. There are 58 and 35 peaks in total
for burst $678$ and $1606$, respectively, although only part of the 
time histories are shown.
\label{2grb-fig}}
\end{figure}

\begin{figure}
\caption{A schematic presentation of a selected peak $C_p(t_p)$
and two neighboring valleys $C_1(t_1)$ and $C_2(t_2)$ in a 
burst time history. The condition 
$C_p - C_{1,2} \geq N_{\rm var}$ $\protect \sqrt{C_p}$ with
$N_{\rm var} \geq 3$ is satisfied, ensuring $C_p$ is a true peak
not a random statistical fluctuation.
The line $B_{1,p,2}$ shows the background level.
The peak fluence is defined approximately as
$S_i = \sum^{t_1}_{t_2} \left [ C(t) - B(t) \right]$
and peak interval $\delta_i = t_{p_{i+1}} - t_{p_i}$.
\label{pfa-fig} }
\end{figure}

\begin{figure}
\caption{The $\chi^2$ probability (filled dots)
for each burst's peak fluences ($a$) 
and peak intervals ($b$) to be log-normally distributed versus the
number of peaks in that burst. Also shown is the probability (triangles)
if peak fluences in each burst are distributed as power-laws ($c$).
Note the logarithmic scale of probability in plot $c$.
Each dot represents one burst (32 total). 
The consistency with log-normal for fluence
and interval distributions and the inconsistency with power-law
for peak fluences are apparent.
\label{chi2prob-fig}}
\end{figure}

\begin{figure}
\caption{The rescaled and sorted peak fluences and
peak intervals from all 32 bursts (dots in all panels)
are compared to both the 
log-normal distribution (dashed line in $a$, hardly visible,
and $c$), 
the Gaussian distribution (dashed line in $b$)
and Poisson interval distribution (dashed line in $d$).
The rescaling is done by $(x_i - <x_i>)/<(x_i - <x_i>)^2>^{1/2}$,
whereas $x_i$ is $\log(S_i)$ in plot $a$, 
$S_i$ in $b$, and $\log(\delta_i)$ in $c$, respectively. 
It is clear that burst distributions are consistent with log-normal
whereas Gaussian hypothesis for fluences and Poisson hypothesis for
intervals are ruled out.
\label{sumks-fig}}
\end{figure}


\begin{thebibliography}{}

\bibitem[Aitchison \& Brown 1957]{ab57} Aitchison, J. \& Brown, 
J.A.C. 1957, The Lognormal Distribution, Cambridge University Press
\bibitem[Cheng et al. 1996]{cheng96} Cheng, B. et al. 1996, Nature, in press
\bibitem[Crow \& Shimizu 1988]{cs88} Crow, E. L. \& Shimizu, K. 1988,
Lognormal Distributions, Marcel Dekker Inc., New York
\bibitem[Fenimore et al. 1995]{fen95} Fenimore, E. E. et al. 1995,
\apj, 448, L101
\bibitem[Hurley et al. 1994]{hur94} Hurley, K. J. et al. 1994, A\&A, 288, L49
\bibitem[Klebesadel 1992]{kle92} Klebesadel, R. W. 1992, in ``Gamma-ray 
Bursts'', eds. Ho, C., Epstein, R., Fenimore, E. E., Cambridge University
Press, P161
\bibitem[Kouveliotou et al. 1993]{k93} Kouveliotou, C. et al. 1993,
\apj, 413, L101
\bibitem[Lindgren 1976]{lin76} Lindgren, B. W. 1976, Statistical Theory,
(3rd ed.; New York: Macmillan)
\bibitem[McBreen et al. 1994]{mc94} McBreen, B. et al. 1994, MNRAS, 271, 662
\bibitem[Meegan et al. 1996]{mee96} Meegan, C. A. et al. 1996, 
\apjs, in press
\bibitem[Mitrofanov et al. 1994]{m94} Mitrofanov, I. G. et al. 1994,
in AIP Conf. Proc. 307, Gamma-Ray Bursts, ed. G.J. Fishman, J.J. 
Brainerd, \& K.C. Hurley (New York:AIP), 187
\bibitem[Norris et al. 1996]{n96} Norris, J. P. et al. 1996, ApJ, 459, 393
\bibitem[Norris et al. 1994]{n94} Norris, J. P. et al. 1994, ApJ, 424, 540

\end{thebibliography}
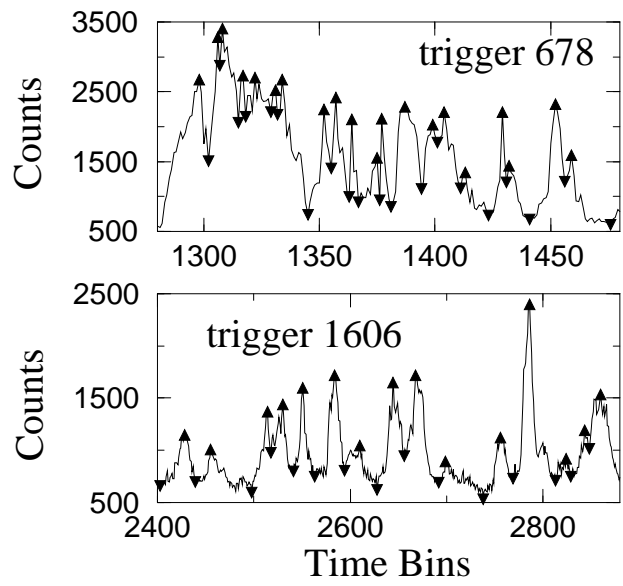
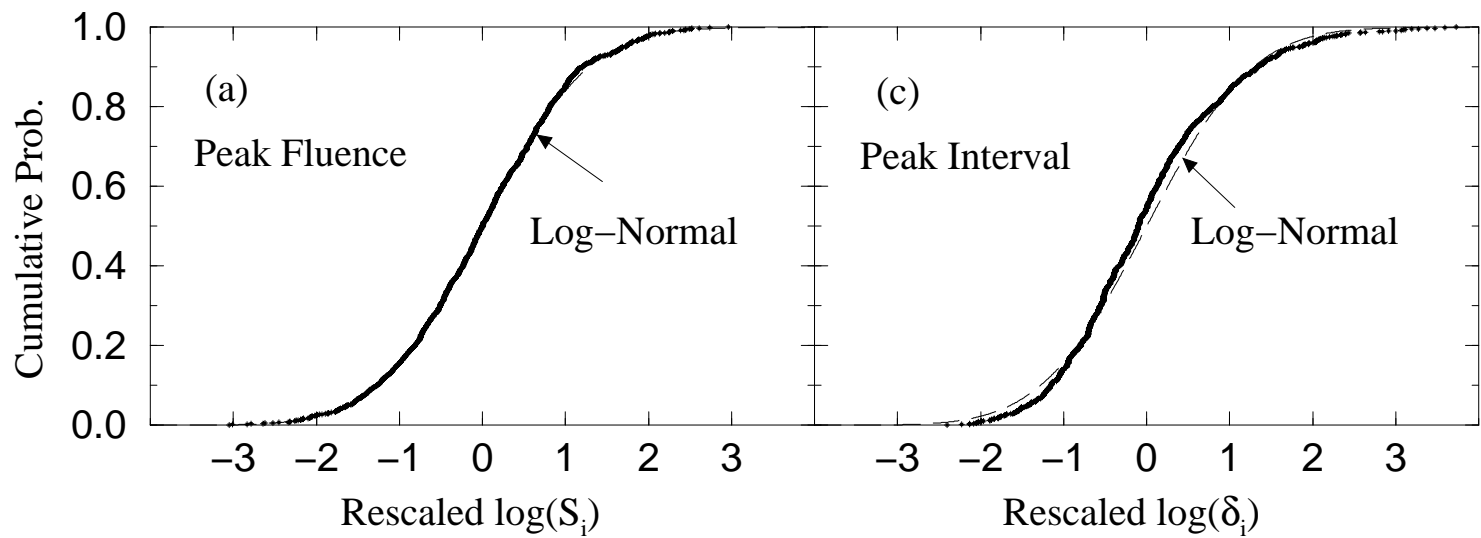
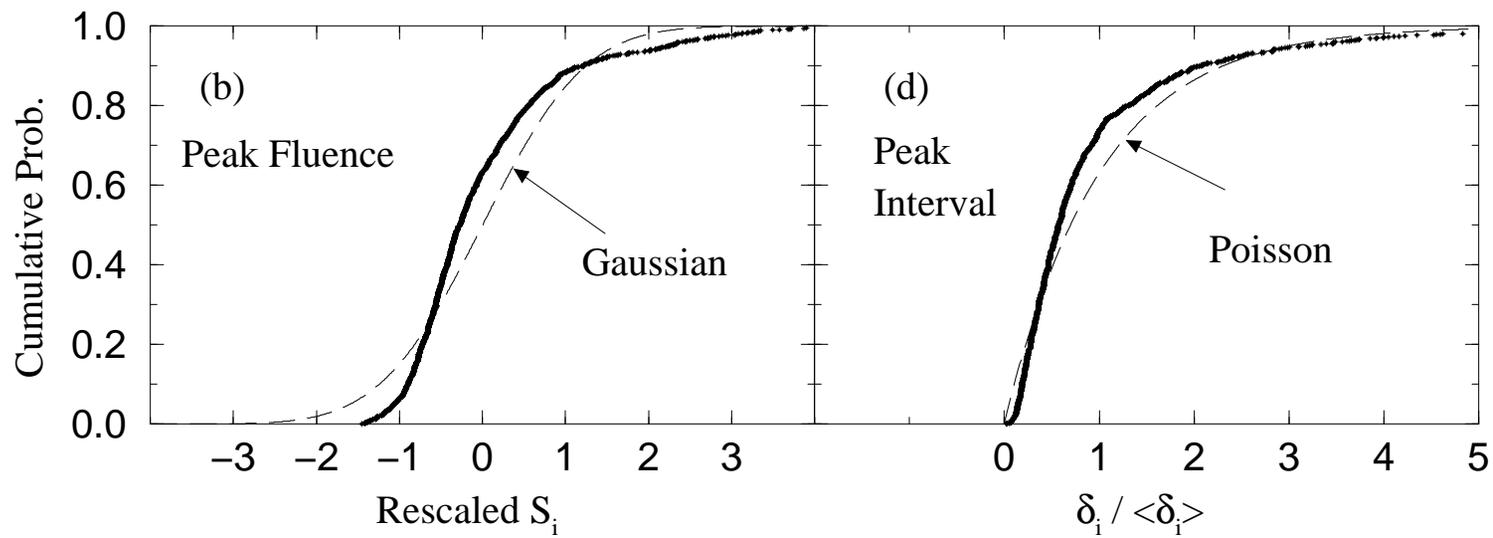
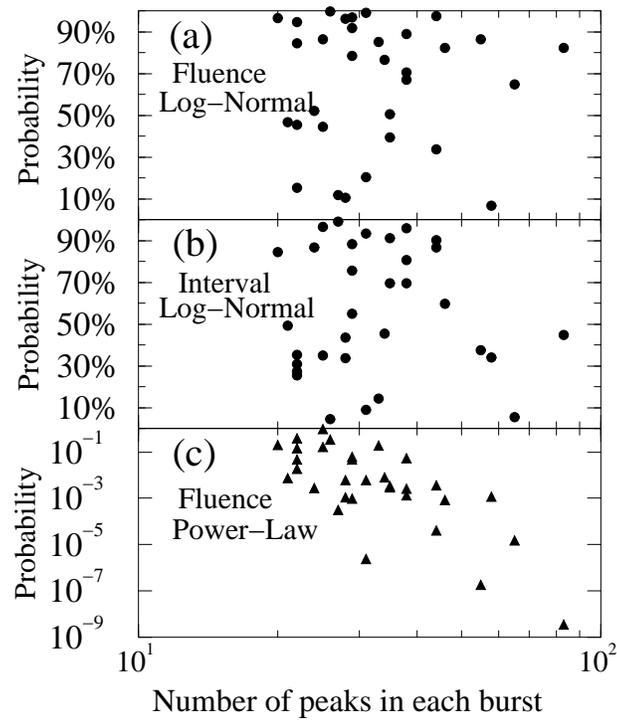
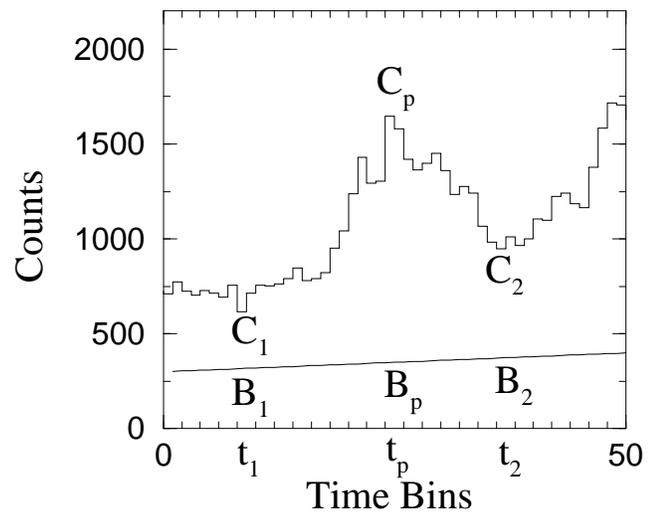
\end{document}